\documentclass[conference]{IEEEtran}
\IEEEoverridecommandlockouts

\usepackage{cite}
\usepackage{amsmath,amssymb,amsfonts}
\usepackage{algorithmic}
\usepackage{graphicx}
\usepackage{textcomp}
\usepackage{xcolor}
\usepackage{comment}
\usepackage{subcaption}
\usepackage{tikz}
\usepackage{amsmath,amssymb,braket}
\usepackage{booktabs}
\usepackage{tabularx}
\usepackage{orcidlink}

\def\BibTeX{{\rm B\kern-.05em{\sc i\kern-.025em b}\kern-.08em
    T\kern-.1667em\lower.7ex\hbox{E}\kern-.125emX}}
    
\begin{document}

\title{Implicit Differentiation for Measurement-Efficient Bilevel Quantum-Classical Optimization\\
}

\author{
    \IEEEauthorblockN{
        Tobias Rohe\IEEEauthorrefmark{1}\IEEEauthorrefmark{4}\orcidlink{0009-0003-3283-0586},
        Markus Baumann\IEEEauthorrefmark{1}\orcidlink{0009-0007-3575-1006},
        Federico Harjes Ruiloba\IEEEauthorrefmark{1}\IEEEauthorrefmark{2}\orcidlink{0009-0002-2283-0921}, 
        Maximilian Zorn\IEEEauthorrefmark{1}\orcidlink{0009-0006-2750-7495}, \\
        Jonas Stein\IEEEauthorrefmark{1}\IEEEauthorrefmark{3}\orcidlink{0000-0001-5727-9151}, and 
        Claudia Linnhoff-Popien\IEEEauthorrefmark{1}\orcidlink{0000-0001-6284-9286}
    }
    \IEEEauthorblockA{
        \IEEEauthorrefmark{1}Institute for Computer Science, LMU Munich, 80538 Munich, Germany\\
    }
    \IEEEauthorblockA{
        \IEEEauthorrefmark{2}relAI – Konrad Zuse School of Excellence in Reliable AI, LMU Munich, 80799 Munich, Germany
    }
    \IEEEauthorblockA{
        \IEEEauthorrefmark{3}Aqarios GmbH, 81677 Munich, Germany
    }
    \IEEEauthorblockA{
        \IEEEauthorrefmark{4}Email: tobias.rohe@ifi.lmu.de
    }
}

\maketitle

\begin{abstract}
Quantum optimization has shown promising results for quadratic unconstrained binary optimization (QUBO) problems. Real-world applications, however, often involve polynomial coefficients that depend on tunable external factors—such as demand forecasts or risk preferences—giving rise to bilevel optimization structures. We show how variational quantum algorithms (VQAs) can efficiently handle such parametric problems, making three contributions. First, we propose a bilevel optimization model for diagonal cost Hamiltonians where coefficients depend on a tunable outer parameter: an outer loop adjusts this parameter—reshaping the cost landscape—while an inner VQA optimizes circuit variables. Second, since derivative-free probing methods incur a multiplicative overhead when each outer evaluation requires a complete inner solve, we develop correlator-reuse implicit differentiation (CR-ID), which obtains outer gradients by reusing quantum measurements already collected during inner energy estimation, requiring essentially no additional circuit executions. Experiments across three coefficient families show that CR-ID consistently improves budget-normalized efficiency by ~4\% in 1D and over 14\% in multi-dimensional settings, showing a significant performance advantage compared to finite-difference methods. Third, we show that this property is architecture-dependent: VQE admits exact reuse gradients, whereas QAOA introduces a state-dependent term that creates a cost–bias trade-off.
\end{abstract}

\begin{IEEEkeywords}
Quantum-classical hybrid optimization, Bilevel optimization, Variational quantum algorithms, Implicit differentiation, QUBO, Max-Cut, Derivative-free optimization
\end{IEEEkeywords}

\section{Introduction}
Variational quantum algorithms (VQAs)—most notably the Quantum Approximate Optimization Algorithm (QAOA)~\cite{farhi2014quantum} and the Variational Quantum Eigensolver (VQE)~\cite{peruzzo2014variational}—are among the leading approaches for near-term quantum optimization~\cite{Cerezo2021VQAReview, amaro2022filtering, Blekos2024QAOAReview}. Most existing work treats the cost Hamiltonian as fixed: given a problem instance, the VQA optimizes circuit parameters to minimize the corresponding energy. In practice, however, the effective cost coefficients often depend on external parameters that are themselves part of the decision process or subject to uncertainty. In portfolio optimization, a risk-aversion parameter directly scales the portfolio variance, fundamentally reshaping optimal allocations as investor preferences change~\cite{buonaiuto2023best}. Supply chain and logistics problems depend on demand forecasts that alter facility and routing costs~\cite{feld2019hybrid}. The time-dependent traveling salesperson problem incorporates a global temporal parameter that determines sequence-dependent traversal costs~\cite{picard1978time}. More generally, any setting in which a combinatorial cost function is parameterized by a continuous control produces a family of problem instances indexed by that control. When such external parameter is treated as a decision variable rather than a fixed constant, VQA optimization becomes a bilevel problem: an outer loop searches over the control parameter, reshaping the cost landscape, while an inner VQA solves the resulting instance for each candidate value. 

This bilevel setting has received great attention in classical optimization~\cite{sinha2017review}; however, little attention in quantum optimization~\cite{deng2020real, fan2021bi, karahalios2025quantum}. 
In standard VQA optimization, measurement cost is already substantial, as gradient estimation via parameter-shift rules requires $O(2p)$ circuit evaluations for $p$ parameters~\cite{schuld2019evaluating}, and achieving precision $\varepsilon$ in expectation values demands $O(\varepsilon^{-2})$ samples~\cite{Cerezo2021VQAReview}. 
When this inner optimization is nested inside an outer loop over $\lambda$---so that the outer objective is accessed only through the converged inner solve, i.e., via the value function $F(\lambda)=\min_{\boldsymbol{\phi}} J(\boldsymbol{\phi},\lambda)$ rather than an end-to-end differentiable computation graph---derivative-free methods for the outer problem require individual inner solves at perturbed parameter values to estimate each outer gradient. 
The result is a multiplicative overhead, as the total measurement budget scales as the product of outer iterations, inner iterations per outer step, and gradient cost per inner iteration.

We address this efficiency gap through the proposed correlator-reuse implicit differentiation (CR-ID) technique, a method that obtains outer gradients at essentially zero additional measurement cost. The key insight follows from the envelope theorem~\cite{milgrom2002envelope}: at inner optimality, the derivative of the optimal value function with respect to outer parameters reduces to a partial derivative of the Hamiltonian expectation—a weighted sum of Pauli correlators that are already measured during inner energy estimation. Rather than probing the outer landscape through repeated inner solves, CR-ID extracts outer gradient information directly from measurements collected in the final inner optimization stage. Importantly, for VQEs with hardware-efficient ansätze, the quantum state at fixed circuit parameters does not depend on the outer parameter, so correlator-reuse yields the exact outer gradient. For QAOA, however, the cost Hamiltonian enters the ansatz through the cost unitary, introducing an additional state-dependence term that creates a cost–bias trade-off: one may either estimate this term at additional measurement cost or accept a biased gradient signal. We characterize this distinction analytically and demonstrate its practical implications. Empirical evaluation across linear, quadratic, and periodic coefficient families shows that CR-ID achieves systematic efficiency gains in measurement-limited regimes.

This paper makes three contributions:
\begin{itemize}
    \item Using weighted Max-Cut as a representative problem, we introduce a bilevel optimization formulation for parametric diagonal cost Hamiltonians with parameter-dependent coefficients.
    \item We empirically demonstrate that CR-ID consistently outperforms finite-difference value probing under matched evaluation budgets: approximately 4\% relative improvement in $AUC_B$ 1D scalar outer control, increasing to over 14\% in multi-dimensional settings where probe overhead compounds.
    \item We characterize CR-ID's architecture dependence both analytically and empirically: VQE admits exact reuse gradients, whereas QAOA introduces a state-dependence term that creates a cost–bias trade-off.
\end{itemize}
The efficiency gain reflects the $\approx 3\times$ evaluation overhead inherent in central finite-difference probing, which CR-ID avoids entirely, resulting in this performance advantage.

\section{Background}\label{background}
\subsection{Maximum Cut Problem}
\label{sec:background_maxcut}

The \emph{Maximum Cut} (Max-Cut) problem is defined on an undirected graph $G = (V, E)$ with vertex set $V$ of cardinality $|V| = n$ and edge set $E$. Each edge $e = (i,j) \in E$ connects two distinct vertices $i, j \in V$ and carries a nonnegative weight $w_{ij} \geq 0$. 

The objective is to partition the vertex set $V$ into two disjoint subsets $S$ and $\bar{S} = V \setminus S$ such that the total weight of edges between the partitions is maximized. We encode each vertex assignment by a binary decision variable $x_i \in \{0,1\}$, where $x_i = 0$ if vertex $i \in S$ and $x_i = 1$ if $i \in \bar{S}$. An edge $e = (i,j)$ is \emph{cut} if and only if $x_i \neq x_j$, which can be expressed algebraically as
\begin{equation}
    \mathbb{I}[x_i \neq x_j] = x_i + x_j - 2x_i x_j.
    \label{eq:edge_cut_indicator}
\end{equation}
The Max-Cut problem can then be formulated as QUBO problem:
\begin{equation}
    \arg\max_{x \in \{0,1\}^n} \sum_{(i,j) \in E} w_{ij} \bigl( x_i + x_j - 2x_i x_j \bigr).
    \label{eq:maxcut_qubo}
\end{equation}

Max-Cut is NP-hard~\cite{Karp1972Reducibility}. A random cut achieves, in expectation, half of the total edge weight, yielding a $1/2$-approximation (derandomizable). The problem is further APX-hard, i.e., it admits no PTAS unless $P=NP$~\cite{PapadimitriouYannakakis1991}; more sharply, it is NP-hard to approximate within $16/17+\varepsilon$~\cite{haastad2001some}. The best known polynomial-time approximation is the Goemans--Williamson algorithm~\cite{goemans1995improved}, which achieves $\alpha_{\mathrm{GW}} \approx 0.87856$---and is optimal assuming the Unique Games Conjecture~\cite{khot2002power}.

\paragraph{Spin formulation.}
To connect the Max-Cut problem to quantum mechanical systems, we map the binary variables
$x_i \in \{0,1\}$ to spin variables $s_i \in \{\pm 1\}$ via the bijection
\begin{equation}
    s_i = 1 - 2x_i \qquad \Leftrightarrow \qquad x_i = \frac{1 - s_i}{2}.
    \label{eq:x_to_s}
\end{equation}
Under this substitution, the edge-cut indicator~\eqref{eq:edge_cut_indicator} becomes
\begin{equation}
    x_i + x_j - 2x_i x_j = \frac{1 - s_i s_j}{2},
    \label{eq:spin_transformation}
\end{equation}
and the Max-Cut objective~\eqref{eq:maxcut_qubo} transforms to
\begin{equation}
    \arg\max_{s \in \{-1,+1\}^n} \sum_{(i,j)\in E} w_{ij}\,\frac{1 - s_i s_j}{2}.
    \label{eq:maxcut_spin}
\end{equation}
Since
$\sum_{(i,j)} w_{ij}\frac{1 - s_i s_j}{2}
= \frac12\sum_{(i,j)} w_{ij} - \frac12\sum_{(i,j)} w_{ij}s_i s_j$,
maximizing~\eqref{eq:maxcut_spin} is equivalent (up to an additive constant) to minimizing an
antiferromagnetic Ising energy, i.e.,
\begin{equation}
    \arg\max_{s}\ \sum_{(i,j)} w_{ij}\frac{1 - s_i s_j}{2}
    \;=\;
    \arg\min_{s}\ \sum_{(i,j)} w_{ij} s_i s_j,
    \label{eq:argmax_argmin}
\end{equation}
establishing a direct correspondence between combinatorial optimization and statistical physics~\cite{lucas2014ising}.

\paragraph{Hamiltonian encoding.}
Let $\{\ket{x}\}_{x\in\{0,1\}^n}$ denote the computational basis.
The Pauli-$Z$ operator on qubit $i$ acts as
\begin{equation}
    Z_i \ket{x} = (-1)^{x_i}\ket{x},
\end{equation}
so that the classical spin variable $s_i \in \{\pm 1\}$ is represented by the $Z$-eigenvalue.
Promoting classical spins to quantum operators via $s_i \mapsto Z_i$ yields the \emph{cost Hamiltonian}
\begin{equation}
    H_C \;=\; \sum_{(i,j)\in E} w_{ij}\, \Pi_{ij},
    \qquad
    \Pi_{ij} \;:=\; \frac{I - Z_i Z_j}{2},
    \label{eq:cost_hamiltonian}
\end{equation}
where $I$ is the identity.
The two-qubit operator $\Pi_{ij}$ is a projector onto the subspace where qubits $i$ and $j$ differ in the
computational basis; equivalently,
\begin{equation}
\begin{aligned}
    \Pi_{ij}
    \;=\;
    \sum_{x:\,x_i\neq x_j} \ket{x}\!\bra{x}, \\
    \qquad\text{so that}\qquad
    \Pi_{ij}\ket{x} = \mathbb{I}[x_i\neq x_j]\ket{x},
    \label{eq:edge_projector}
\end{aligned}
\end{equation}
and in particular $\bra{x}\Pi_{ij}\ket{x}=\mathbb{I}[x_i\neq x_j]$.
Consequently, the classical cut value of an assignment $x$ is recovered as the diagonal matrix element
\begin{equation}
    \bra{x}H_C\ket{x} \;=\; \sum_{(i,j)\in E} w_{ij}\,\mathbb{I}[x_i\neq x_j].
    \label{eq:hamiltonian_eigenvalue}
\end{equation}

Because each term $\Pi_{ij}$ is a polynomial in Pauli-$Z$ operators, it is diagonal in the computational ($Z$) basis.
Therefore, all edge terms commute,
$[\Pi_{ij},\Pi_{k\ell}]=0$ for all $(i,j),(k,\ell)\in E$,
and $H_C$ can be estimated from a \emph{single} measurement setting: measuring \emph{all} qubits in the computational
($Z$) basis.
For a general $n$-qubit state $\rho$, this measurement induces the distribution
$p(x)=\bra{x}\rho\ket{x}$, under which the probability that edge $(i,j)$ is cut equals
\begin{equation}
    p_{ij}
    \;:=\;
    \mathrm{Tr}(\rho\,\Pi_{ij})
    \;=\;
    \sum_{x:\,x_i\neq x_j} p(x)
    \;=\;
    \frac{1-\langle Z_i Z_j\rangle_\rho}{2}.
    \label{eq:edge_cut_probability}
\end{equation}
The expected cut value then decomposes edge-wise as
\begin{equation}
    \mathrm{Tr}(\rho\,H_C) \;=\; \sum_{(i,j)\in E} w_{ij}\,p_{ij}.
    \label{eq:edge_decomposition}
\end{equation}
This decomposition enables \emph{correlator-reuse}: each $Z$-basis shot returns a bitstring $x$, from which all
indicators $\mathbb{I}[x_i\neq x_j]$ (and thus all $\{p_{ij}\}_{(i,j)\in E}$) can be computed by classical
post-processing, simultaneously yielding an estimate of $\mathrm{Tr}(\rho H_C)$.

\subsection{Variational Quantum Eigensolver}
\label{sec:background_vqe}

The Variational Quantum Eigensolver (VQE) is a hybrid quantum-classical algorithm designed to estimate the ground state energy $E_0$ of a target Hamiltonian $H$~\cite{peruzzo2014variational}. It relies on the \emph{Rayleigh--Ritz variational principle}~\cite{rayleigh1870finding, Ritz1908}, which states that for any normalized trial state $\ket{\psi(\boldsymbol{\theta})}$, the expectation value of the Hamiltonian provides an upper bound on the ground state energy:
\begin{equation}
    E(\boldsymbol{\theta}) = \bra{\psi(\boldsymbol{\theta})} H \ket{\psi(\boldsymbol{\theta})} \geq E_0,
    \label{eq:variational_principle}
\end{equation}
where $\boldsymbol{\theta} \in \mathbb{R}^p$ denotes a vector of $p$ classical parameters. The VQE algorithm seeks to minimize $E(\boldsymbol{\theta})$ over the parameter space to approximate $E_0$ and acquire information on the corresponding ground state $\ket{\psi(\boldsymbol{\theta}^*)}$. It operates as an iterative quantum-classical feedback loop comprising three stages:

\paragraph{State preparation.}
A parameterized quantum circuit $U(\boldsymbol{\theta})$, termed the \emph{ansatz}, prepares the trial state from a fixed initial state (typically the computational basis state $\ket{0}^{\otimes n}$):
\begin{equation}
    \ket{\psi(\boldsymbol{\theta})} = U(\boldsymbol{\theta}) \ket{0}^{\otimes n}.
    \label{eq:ansatz_state}
\end{equation}

\paragraph{Measurement.}
The Hamiltonian is decomposed into a weighted sum of Pauli strings,
\begin{equation}
    H = \sum_{k=1}^{K} \alpha_k P_k, \qquad P_k \in \{I, X, Y, Z\}^{\otimes n},
    \label{eq:hamiltonian_decomposition}
\end{equation}
where each $P_k$ is a tensor product of single-qubit Pauli operators and $\alpha_k \in \mathbb{R}$ are the corresponding coefficients. The expectation value is then estimated as
\begin{equation}
    E(\boldsymbol{\theta}) = \sum_{k=1}^{K} \alpha_k \bra{\psi(\boldsymbol{\theta})} P_k \ket{\psi(\boldsymbol{\theta})},
    \label{eq:energy_estimation}
\end{equation}
where each term $\langle P_k \rangle$ is obtained by repeated state preparation and measurement in the appropriate Pauli basis. For the Max-Cut cost Hamiltonian~\eqref{eq:cost_hamiltonian}, every term is a Pauli-$Z$ string ($P_k \in \{I, Z\}^{\otimes n}$). Hence, measuring all qubits once in the computational ($Z$) basis yields bitstrings $x$, from which all edge correlators $\langle Z_i Z_j \rangle$ are obtained by classical post-processing and averaging over shots.

\paragraph{Classical optimization.}
A classical optimizer updates the parameters $\boldsymbol{\theta}$ to minimize $E(\boldsymbol{\theta})$. For gradient-based methods, parameters are updated according to
\begin{equation}
    \boldsymbol{\theta}^{(t+1)} = \boldsymbol{\theta}^{(t)} - \eta \nabla_{\boldsymbol{\theta}} E(\boldsymbol{\theta}^{(t)}),
    \label{eq:gradient_update}
\end{equation}
where gradients can be computed via the parameter-shift rule~\cite{schuld2019evaluating} or estimated using gradient-free techniques such as SPSA~\cite{spall2002multivariate} when shot noise dominates. The optimization terminates when a convergence criterion is satisfied or a computational budget is exhausted.

It is important to note that VQE is fundamentally a \emph{heuristic} algorithm with known challenges, including barren plateaus~\cite{mcclean2018barren}. Nevertheless, due to its shallow circuit requirements, its robustness to certain error types, as well as several adaptations, the VQE is a leading candidate in optimization and simulation tasks~\cite{Cerezo2021VQAReview, kolotouros2022evolving}.

\subsection{Quantum Approximate Optimization Algorithm}
\label{sec:background_qaoa}
The Quantum Approximate Optimization Algorithm (QAOA) is a structured ansatz within the variational framework, specifically designed to approximate solutions of combinatorial optimization problems~\cite{farhi2014quantum}. While the VQE accommodates arbitrary parameterized circuits, the QAOA constructs the trial state through a physically motivated procedure derived from the Trotterization of adiabatic quantum computation, making it a problem-specific subclass of variational quantum algorithms.

The algorithm employs an alternating operator ansatz with two non-commuting Hamiltonians: the cost Hamiltonian $H_C$ encoding the optimization objective~\eqref{eq:cost_hamiltonian} and the mixer Hamiltonian $H_M = \sum_{i=1}^{n} X_i$ inducing transitions between basis states. The QAOA state at depth/layer $p$ is constructed as
\begin{equation}
    \ket{\psi(\boldsymbol{\gamma}, \boldsymbol{\beta})} = \prod_{\ell=1}^{p} e^{-i \beta_\ell H_M} e^{-i \gamma_\ell H_C} \ket{+}^{\otimes n},
    \label{eq:qaoa_state}
\end{equation}
where $\boldsymbol{\gamma}, \boldsymbol{\beta} \in \mathbb{R}^p$ are variational parameters optimized to maximize $\langle H_C \rangle$. For Max-Cut, the phase-separation unitary decomposes into commuting two-qubit gates $e^{-i \gamma w_{ij} (I - Z_i Z_j)/2}$ on each edge, while the mixing unitary factorizes into single-qubit $X$-rotations.

A property critical to this work is that $H_C$ appears both in the measurement observable and in state preparation through $e^{-i \gamma_\ell H_C}$. Consequently, when edge weights depend on an external parameter $\lambda$, the prepared state $\ket{\psi(\boldsymbol{\gamma}, \boldsymbol{\beta}; \lambda)}$ inherits this dependence (see Sec.~\ref{sec:architecture})---creating a state-dependence term absent in VQE that we analyze in subsequent sections.

\section{Related Work}
\label{sec:related_work}
Implicit differentiation is well established in classical machine learning for bilevel problems such as hyperparameter optimization and meta-learning, where it enables efficient hypergradient computation without unrolling the inner optimization trajectory~\cite{lorraine2020optimizing, rajeswaran2019meta, franceschi2018bilevel}.
In the domain of variational quantum algorithms, by contrast, gradient computation has focused almost exclusively on circuit parameters via the parameter-shift rule, which obtains exact gradients by evaluating circuits at shifted parameter values~\cite{mitarai2018quantum, schuld2019evaluating}. Subsequent work has generalized these rules to broader gate families and multi-qubit evolutions~\cite{wierichs2022general, banchi2021measuring}.
The application of implicit differentiation to VQAs was introduced by Ahmed et al.~\cite{ahmed2023implicit}, who showed that gradients of observables evaluated on variationally obtained ground states can be computed via the implicit function theorem without unrolling the inner optimization. Their work demonstrated applications to generalized susceptibilities in condensed matter physics and hyperparameter tuning in quantum classifiers, establishing the theoretical foundation for differentiating through VQA solutions. 
The present work extends this framework to parametric combinatorial optimization, where the cost Hamiltonian coefficients depend on an external control parameter and the outer objective is the value function of an inner VQA solve—inducing an explicit bilevel structure. For diagonal cost Hamiltonians such as Max-Cut, we show that the envelope theorem connects naturally to the measurement structure, enabling what we term CR-ID. We further analyze how this property depends on the choice of VQA architecture.

\section{Problem Formulation}
\label{sec:problem_formulation}
We consider weighted Max-Cut instances whose edge couplings vary smoothly with an external control parameter $\lambda \in [\lambda_{\min}, \lambda_{\max}]$. This parametric structure captures scenarios where costs depend on tunable external factors that shape the problem instance itself. For each fixed $\lambda$, a VQA optimizes circuit parameters to solve the resulting instance; when $\lambda$ is also a decision variable, this naturally induces a bilevel optimization problem.

\subsection{Parametric Hamiltonian Family}

Let $G = (V, E)$ be an undirected graph with $|V| = n$ vertices. For each edge $e = (i,j) \in E$, we use the standard Max-Cut projector
\begin{equation}
\Pi_e := \frac{1}{2}(I - Z_i Z_j),
\label{eq:projector}
\end{equation}
which returns the measurement outcome $1$ when qubits $i$ and $j$ differ in the computational basis and $0$ otherwise. The parametric cost Hamiltonian is then
\begin{equation}
H_C(\lambda) = \sum_{e \in E} w_e(\lambda)\, \Pi_e,
\label{eq:hamiltonian}
\end{equation}
where each coupling $w_e(\lambda) > 0$ is a continuous function of the control parameter. This positivity constraint ensures that the problem remains a valid weighted cut for all $\lambda$ in the domain. Throughout, we consider both a scalar control $\lambda \in \mathbb{R}$ (\emph{1D setting}) and an edge-wise control vector $\boldsymbol{\lambda} = (\lambda_e)_{e \in E} \in \mathbb{R}^{|E|}$ (\emph{multi-dimensional setting}), where each edge weight depends on its own control parameter $w_e(\lambda_e)$.

\subsection{Bilevel Optimization Structure}

Given a VQA with circuit parameters $\boldsymbol{\phi}$, the inner objective at a fixed $\lambda$ is
\begin{equation}
J(\boldsymbol{\phi}, \lambda) = \langle \psi(\boldsymbol{\phi}) | H_C(\lambda) | \psi(\boldsymbol{\phi}) \rangle = \sum_{e \in E} w_e(\lambda)\, p_e(\boldsymbol{\phi}),
\label{eq:inner_objective}
\end{equation}
where $p_e(\boldsymbol{\phi}) = \mathrm{Tr}(\rho(\boldsymbol{\phi}) \Pi_e)$ is the probability that edge $e$ is cut under the circuit-induced distribution. The \emph{value function}
\begin{equation}
F(\lambda) = \max_{\boldsymbol{\phi}} J(\boldsymbol{\phi}, \lambda)
\label{eq:value_function}
\end{equation}
captures the best achievable objective after inner optimization at each $\lambda$, and the outer problem seeks to maximize $F(\lambda)$ over the admissible control interval. This bilevel structure—where each outer evaluation requires an inner VQA solve—is the source of the cost considerations analyzed in the following section.

\section{Correlator-Reuse Implicit Differentiation}
\label{sec:method}
This section presents the core methodological contribution: a probe-free outer update signal for bilevel tuning of parametric Max-Cut Hamiltonians. While we use parametric Max-Cut Hamiltonians as a running example, the CR-ID construction extends directly to any diagonal cost Hamiltonian with $\lambda$-dependent coefficients, since the objective is always a weighted sum of correlators obtainable from the same energy-evaluation data. We first establish the cost model that motivates avoiding value-function probes, then derive the CR-ID signal, and finally analyze its architecture dependence.

\subsection{Cost Model and Budget Primitive}
\label{sec:cost_model}
In bilevel VQA tuning, the outer objective is the value function $F(\lambda) = \max_{\boldsymbol{\phi}} J(\boldsymbol{\phi}, \lambda)$, where each evaluation $\widehat{F}(\lambda)$ requires an inner VQA solve. We measure the execution cost in \emph{energy evaluations}---the number of objective estimates $\widehat{J}(\boldsymbol{\phi}, \lambda)$ performed---because this captures the dominant unit of algorithmic work shared across evaluation settings.

For Max-Cut Hamiltonians, one energy evaluation consists of estimating
\begin{equation}
J(\boldsymbol{\phi}, \lambda) = \sum_{e \in E} w_e(\lambda) \, p_e(\boldsymbol{\phi}),
\label{eq:J_decomposition}
\end{equation}
where $p_e(\boldsymbol{\phi}) = \mathrm{Tr}(\rho(\boldsymbol{\phi}) \Pi_e)$ is the probability that edge $e$ is cut under the circuit-induced distribution. Because all Max-Cut terms commute and are diagonal in the computational basis, a single batch of $Z$-basis measurements yields estimates for all edge correlators simultaneously.

Derivative-free outer methods that construct directions from value-function probes face a structural overhead. If an outer update requires $M$ probes $\{\widehat{F}(\lambda + \delta_i)\}_{i=1}^M$, and each inner solve costs $N_{\mathrm{inner}}$ energy evaluations, the cost per outer step scales as
\begin{equation}
C_{\mathrm{outer}} \approx M \cdot N_{\mathrm{inner}}.
\label{eq:probe_overhead}
\end{equation}
Here $M$ denotes the number of full outer objective evaluations (each requiring an inner solve) per outer update.
Central finite differences use $M\!\approx\!3$ evaluations ($\lambda$, $\lambda\!+\!\epsilon$, $\lambda\!-\!\epsilon$), while SPSA uses $M\!\approx\!2$.
Thus, under a fixed evaluation budget $B$, the number of outer updates scales as $\approx B/M$.

\subsection{Response Framework}
\label{sec:response_framework}

To enable controlled benchmarking across different parametric regimes, we separate the scale and shape of weight variation. First,
we map $\lambda$ to a normalized coordinate $x(\lambda) = 2(\lambda - \lambda_{\min})/(\lambda_{\max} - \lambda_{\min}) - 1 \in [-1, 1]$. Each edge weight then follows the template
\begin{equation}
w_e(\lambda) = \bar w_e + A_e \cdot f_e\!\bigl(x(\lambda)\bigr),
\label{eq:response_template}
\end{equation}
where $\bar w_e > 0$ is a baseline weight, $A_e \ge 0$ controls the variation amplitude, and $f_e : [-1, 1] \to \mathbb{R}$ is a shape function satisfying
$\mathbb{E}[f_e] = 0$ and $\mathbb{E}[f_e^2] = 1$ under the uniform distribution on $[-1, 1]$, so that the orders of magnitude are comparable.
In our benchmark instances, we draw baseline weights $\bar w_e \sim \mathrm{Unif}[2,3]$ and amplitudes $A_e \sim \mathrm{Unif}[0.3,0.8]$, which ensures $w_e(\lambda)>0$ throughout $\lambda\in[\lambda_{\min},\lambda_{\max}]$.
The prefactors in the definitions below are chosen exactly to enforce the normalization $\mathbb{E}[f_e]=0$ and $\mathbb{E}[f_e^2]=1$ for $x\sim\mathrm{Unif}[-1,1]$, so that $A_e$ has a consistent interpretation as the variation scale across families.

We consider three canonical response families with increasing complexity:
\begin{itemize}
\item \textbf{Linear:} $f^{\mathrm{lin}}_e(x) = \sqrt{3}\, s_e x$, where $s_e \in \{\pm 1\}$ determines the trend direction.
\item \textbf{Quadratic:} $f^{\mathrm{quad}}_e(x) = \sqrt{45/4}\, s_e (x^2 - 1/3)$, introducing curvature while remaining smooth.
\item \textbf{Periodic:} $f^{\mathrm{per}}_e(x) = \sqrt{2}\cos(\pi k_e x + \varphi_e)$, with frequency $k_e \sim \mathrm{Unif}\{1,\ldots,K\}$ and phase $\varphi_e \sim \mathrm{Unif}[0,2\pi)$.
\end{itemize}
The periodic family includes a difficulty parameter $K$ that controls the maximum oscillation frequency; larger $K$ produces more frequent changes in the identity of the optimal bitstring across $\lambda$,
creating a natural stress test for outer-loop methods.

\subsection{Envelope Principle and Implicit Differentiation}
\label{sec:envelope}
The envelope theorem is the key insight for avoiding probe-based overhead in the \emph{outer} variable $\lambda$.
Let $\boldsymbol{\phi}^*(\lambda)$ denote an inner maximizer (e.g., obtained by standard VQA training in $\phi$)
satisfying the stationarity condition $\nabla_{\boldsymbol{\phi}} J(\boldsymbol{\phi}^*(\lambda), \lambda)=0$.
Defining $F(\lambda):=J(\boldsymbol{\phi}^*(\lambda),\lambda)$, the chain rule gives
\begin{equation}
\begin{aligned}
&\frac{d}{d\lambda}F(\lambda)= \\
&\frac{\partial}{\partial\lambda}J(\phi^\ast(\lambda),\lambda)
+
\left\langle \nabla_{\phi}J(\phi^\ast(\lambda),\lambda),\frac{d\phi^\ast(\lambda)}{d\lambda}\right\rangle.
\end{aligned}
\label{eq:method_chain_rule}
\end{equation}
At inner stationarity, the second term vanishes, yielding the envelope identity
\begin{equation}
\frac{d}{d\lambda}F(\lambda)=\frac{\partial}{\partial\lambda}J(\phi^\ast(\lambda),\lambda).
\label{eq:method_envelope_identity}
\end{equation}
Thus, the outer derivative depends only on the \emph{partial} derivative w.r.t.\ $\lambda$ and does not require
differentiating the optimizer map $\boldsymbol{\phi}^*(\lambda)$; in our setting this partial derivative is obtained
by reweighting correlators already estimated during inner energy evaluation.

\subsection{Correlator-Reuse for Parametric Max-Cut}
\label{sec:correlator_reuse}

We now specialize the envelope identity to parametric Max-Cut. From Eq.~\eqref{eq:J_decomposition}, differentiating with respect to $\lambda$ at fixed $\boldsymbol{\phi}$ yields
\begin{equation}
\frac{\partial}{\partial \lambda}J(\boldsymbol{\phi}, \lambda)
= \sum_{e \in E} \frac{dw_e(\lambda)}{d\lambda} \, p_e(\boldsymbol{\phi}).
\label{eq:CR_signal}
\end{equation}
Correlator-reuse assumes access to the weight sensitivities $dw_e(\lambda)/\allowbreak d\lambda$ (a \emph{white-box} assumption). For the response families in Section~\ref{sec:response_framework}, these derivatives are analytic. If only black-box access to the weights is available but $w_e(\lambda)$ can be queried, one may instead use a purely classical finite-difference approximation, e.g.,
\begin{equation}
\frac{dw_e(\lambda)}{d\lambda}\approx \frac{w_e(\lambda+\varepsilon)-w_e(\lambda-\varepsilon)}{2\varepsilon},
\end{equation}
which still does not require any additional quantum measurements.

The edge cut probabilities $p_e(\boldsymbol{\phi})$ are precisely the quantities estimated during energy evaluation from the same batch of $Z$-basis shots. Therefore, the CR-ID outer signal
\begin{equation}
\widehat{g}_{\mathrm{CR}}(\lambda) := \sum_{e \in E} \frac{dw_e(\lambda)}{d\lambda} \, \widehat{p}_e
\label{eq:CR_estimator}
\end{equation}
is available at essentially zero additional measurement cost once $\widehat{F}(\lambda)$ has been evaluated.

Under shot-based evaluation with $S$ samples, let $\{z^{(s)}\}_{s=1}^S$ be i.i.d.\ bitstrings from measuring the circuit state at parameters $\boldsymbol{\phi}$ in the computational basis. Defining
\begin{equation}
\widehat{p}_e := \frac{1}{S}\sum_{s=1}^S \mathbb{I}\{z_i^{(s)} \neq z_j^{(s)}\}.
\end{equation}
By construction, $\mathbb{E}[\widehat{p}_e]=p_e(\boldsymbol{\phi})$, and by linearity of expectation,
\begin{equation}
\begin{aligned}
\mathbb{E}\!\left[\widehat{g}_{\mathrm{CR}}(\lambda)\right]
= \sum_{e\in E}\frac{dw_e(\lambda)}{d\lambda}\,\mathbb{E}[\widehat{p}_e] \\
= \sum_{e\in E}\frac{dw_e(\lambda)}{d\lambda}\,p_e(\boldsymbol{\phi})
= \frac{\partial}{\partial\lambda}J(\boldsymbol{\phi},\lambda),
\end{aligned}
\end{equation}
i.e., $\widehat{g}_{\mathrm{CR}}$ is an unbiased estimator of the partial derivative at fixed $\boldsymbol{\phi}$. Moreover, since $\widehat{p}_e$ is a sample mean of Bernoulli indicators, $\mathrm{Var}(\widehat{p}_e)=p_e(1-p_e)/S\le 1/(4S)$, implying that the noise of $\widehat{g}_{\mathrm{CR}}$ scales as $O(1/S)$ and is governed by the magnitudes of the sensitivity weights $dw_e(\lambda)/d\lambda$.

\subsection{Architecture Dependence: VQE versus QAOA}
\label{sec:architecture}

The ``free'' nature of Eq.~\eqref{eq:CR_signal} depends critically on how $\lambda$ enters the computation.

\paragraph{VQE: Clean reuse.} In VQE, the quantum state $\rho(\boldsymbol{\theta})$ depends only on circuit parameters $\boldsymbol{\theta}$; the outer variable $\lambda$ affects only the Hamiltonian coefficients. Therefore, $\partial_\lambda \rho(\boldsymbol{\theta}) = 0$ when $\boldsymbol{\theta}$ is held fixed, and Eq.~\eqref{eq:CR_signal} is the \emph{exact} partial derivative. Combined with the envelope identity, CR-ID provides an unbiased outer gradient estimate at inner stationarity with no additional quantum cost.

\paragraph{QAOA: State-dependence term.} In QAOA, $\lambda$ enters the state preparation through the cost unitary $e^{-i\gamma_\ell H_C(\lambda)}$. The objective
$J(\boldsymbol{\gamma}, \boldsymbol{\beta}, \lambda) = \mathrm{Tr}(\rho(\boldsymbol{\gamma}, \boldsymbol{\beta}, \lambda)\, H_C(\lambda))$
therefore depends on $\lambda$ both explicitly (through coefficients) and implicitly (through the state). Differentiating yields
\begin{equation}
\begin{aligned}
&\frac{\partial}{\partial \lambda}J(\boldsymbol{\gamma}, \boldsymbol{\beta}, \lambda) \\
&= \underbrace{\sum_{e} \frac{dw_e(\lambda)}{d\lambda}\, p_e(\boldsymbol{\gamma}, \boldsymbol{\beta}, \lambda)}_{\text{explicit term}}
+ \underbrace{\sum_{e} w_e(\lambda)\, \frac{\partial p_e}{\partial \lambda}(\boldsymbol{\gamma}, \boldsymbol{\beta}, \lambda)}_{\text{state-dependence term}}.
\end{aligned}
\label{eq:QAOA_decomp}
\end{equation}
The first term has the same correlator-reuse form as VQE. The second term captures how the measurement distribution changes when $\lambda$ changes at fixed circuit parameters $(\boldsymbol{\gamma}, \boldsymbol{\beta})$, and it cannot generally be computed from standard energy-evaluation data. Estimating it requires additional circuit evaluations (e.g., finite differences in $\lambda$ at fixed $(\boldsymbol{\gamma}, \boldsymbol{\beta})$) or specialized gradient circuits, reintroducing cost overhead. This creates a cost--bias trade-off for QAOA: using only the explicit term (``reuse-only'') is cheap but biased, while estimating the full derivative removes bias but incurs additional measurements.

\subsection{Baseline: FD-Style Value Probing}
\label{sec:baseline}

We compare CR-ID against outer methods that estimate directions by probing the value function at perturbed controls. The representative baseline uses central finite differences:
\begin{equation}
\widehat{g}_t^{\mathrm{FD}} := \frac{\widehat{F}(\lambda_t + c_t) - \widehat{F}(\lambda_t - c_t)}{2c_t},
\label{eq:FD_baseline}
\end{equation}
where each $\widehat{F}(\cdot)$ requires an inner VQA solve at the corresponding $\lambda$ value. This baseline exposes the structural overhead of probe-based methods and supports fair matched-budget comparisons: an \emph{outer iteration} is not equally expensive across methods. In particular, FD-style outer updates require multiple objective evaluations (e.g., two evaluations for the central difference probe, and potentially more depending on how many directions/components are probed). If we compared methods by outer iteration count, we would implicitly grant FD baselines a larger evaluation budget. We therefore evaluate all methods against \emph{cumulative energy evaluations} (i.e., the total number of objective/energy calls, equivalently inner VQA solves) rather than outer iteration count.

\section{Experimental Set}
\label{sec:experimental_setup}

\subsection{Evaluation Protocol}
\label{sec:eval_protocol}
All methods are compared under a \emph{matched evaluation budget} $B$ measured in cumulative energy evaluations, representing the number of objective estimates $\widehat{J}(\boldsymbol{\phi}, \lambda)$ performed across all inner solves. This currency makes bilevel overhead explicit.

The performance is tracked via the \emph{best-so-far trajectory}
\begin{equation}
y(b) := \max_{\text{evals} \leq b} \widehat{F}(\lambda),
\label{eq:best_so_far}
\end{equation}
which records the highest value-function estimate observed up to that point $b$. Crucially, this includes all evaluated candidates, including probe points $\lambda_t \pm c_t$ for the FD baseline. For instance-comparison, we normalize by the classical optimum $J^* = \max_{\lambda} \max_{x \in \{0,1\}^n} \allowbreak \langle x | H_C(\lambda) | x \rangle$ and summarize budget efficiency via the area under the curve:
\begin{equation}
\mathrm{AUC}_B := \frac{1}{B} \int_0^B \frac{y(b)}{J^*} \, db.
\label{eq:auc_metric}
\end{equation}
This metric captures \emph{progress per budget}, and therefore, how quickly methods reach high-quality solutions, rather than only the final value at budget exhaustion.

\subsection{Benchmark Instances}
\label{sec:benchmark_instances}

We generate parametric Max-Cut instances on Erd\H{o}s--R\'{e}nyi\\
graphs~\cite{erdos1960evolution} $G(n,p)$ with $n \in \{10,12,14\}$ and edge probabilities $p \in \allowbreak \{0.25, 0.35,\allowbreak 0.45, 0.55, 0.65\}$.
We use 20 instances per $(n,p)$ setting, for a total of $N=300$ instances per experiment. For each graph, edge weights follow the response framework of Section~\ref{sec:response_framework} with baseline weights $\bar{w}_e \sim \mathrm{Unif}[2,3]$ and amplitude $A_e \sim \mathrm{Unif}[0.3,\allowbreak 0.8]$, which ensures $w_e(\lambda) > 0$ throughout $\lambda \in [-5,5]$.

We evaluate all three response families with the periodic family serving as the primary stress test due to its controllable difficulty. Increasing the frequency parameter $K$ (we use $K=6$ unless stated otherwise) produces more frequent changes in the optimal bitstring across $\lambda$, creating increasingly challenging outer landscapes for both gradient-based and probe-based methods.

\subsection{Methods and Fairness Rules}
\label{sec:methods_compared}

We compare two outer-loop strategies:
\begin{itemize}
\item \textbf{CR-ID:} Correlator-reuse implicit differentiation using Eq. \eqref{eq:CR_estimator}. Each outer step requires only the center evaluation $\widehat{F}(\lambda_t)$; the outer signal is computed from the same correlators at zero additional cost.
\item \textbf{FD-Baseline:} Central finite differences on the value function \eqref{eq:FD_baseline} with perturbation $c_t$. Each outer step evaluates $\widehat{F}(\lambda_t)$, $\widehat{F}(\lambda_t + c_t)$, and $\widehat{F}(\lambda_t - c_t)$, yielding $M = 3$ inner solves per update.
\end{itemize}

Both methods use identical inner solvers (SPSA-based VQE with hardware-efficient ansatz), identical outer step-size schedules, and identical warm-start initialization.
The only difference is how the outer direction is obtained.
All comparisons are \emph{paired per instance}: for each problem instance, CR-ID and the FD baseline are run on the same instance under the same protocol. The evaluation budget $B=1830$ is determined by the FD baseline's requirements; $T=30$ outer steps with $I=20$ SPSA inner iterations, $B=T(3\cdot I+1)=30\cdot(3\cdot 20 + 1)=1830$ (three objective evaluations per SPSA iteration plus one final evaluation per outer step). CR-ID then operates under this same matched budget.

Table~\ref{tab:exp_defaults} summarizes the experimental defaults.

\begin{table}[t]
\centering
\caption{Experimental defaults.}
\label{tab:exp_defaults}

\scriptsize
\setlength{\tabcolsep}{4pt}
\renewcommand{\arraystretch}{1.0}

\begin{tabular}{@{}p{0.34\columnwidth}p{0.62\columnwidth}@{}}
\toprule
\textbf{Setting} & \textbf{Value} \\
\midrule
Graph ensemble
& Erd\H{o}s--R\'{e}nyi $G(n,p)$; $n\in\{10,12,14\}$; $p\in\{0.25,0.35,0.45,0.55,0.65\}$ \\
Outer steps
& $30$ \\
Inner iterations
& $10$ (Exp.~9); $20$ (Exp.~10,12) \\
Matched evaluation budget
& $B = 1830$ \\
Ansatz depth
& VQE depth $L_{\mathrm{VQE}}=2$; QAOA depth $p_{\mathrm{QAOA}}=3$ (Exp.~10) \\
Energy eval cost
& one commuting $Z$-basis shot (shots\_per\_eval$=1$) \\
Readout evaluation
& best-of-$32$ samples \\
Tail threshold
& $\varepsilon=0.1$, $p_{\mathrm{target}}=0.99$ \\
Instances
& $N=300$ per experiment \\
\bottomrule
\end{tabular}
\end{table}

\subsection{Metrics}
\label{sec:metrics}
For each comparison, we report (mean $\pm$ std over $N=300$):
\begin{itemize}
\item Best-so-far normalized objective at budget, $y(B)/J^*$;
\item Budget efficiency AUC$_B$ from Eq.~\eqref{eq:auc_metric};
\item Readout best normalized objective at budget (best-of-32 samples);
\item Tail hit-rate $p_{\mathrm{good}}$ at threshold $\varepsilon=0.1$.
\end{itemize}
For readout-facing experiments, we additionally report best-of-$S$ sampled cut values and per-shot near-optimal hit rates to bridge expectation-level optimization to operational solution quality.

\section{Results}
\label{sec:results}
We evaluate CR-ID against the FD baseline across three axes: (1) systematic budget efficiency across response families (linear, quadratic, periodic) in 1D outer control; (2) scaling to multi-dimensional outer control; and (3) an architecture comparison between VQE and QAOA.

\subsection{Visualizing the Bilevel Optimization Process}
\label{sec:results_visualization}
\begin{figure}[t]
\centering

\newcommand{\panelbox}[2]{%
\begin{minipage}[t]{0.49\columnwidth}
  \centering
  \textbf{(#1)}\par\vspace{0.25em}
  \includegraphics[width=\linewidth,height=0.95\linewidth,keepaspectratio]{figures/#2}
\end{minipage}%
}

\panelbox{a}{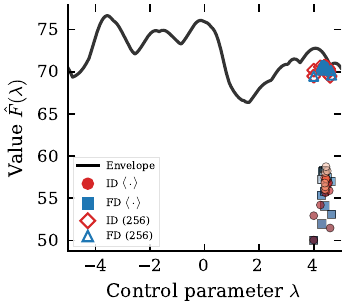}\hfill
\panelbox{b}{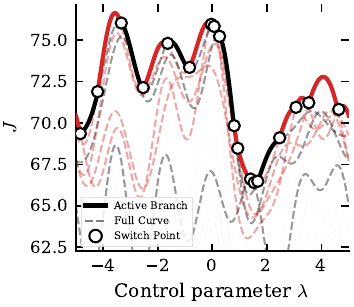}

\caption{\textbf{Bilevel optimization diagnostics VQE on periodic family.}
(a)~Classical envelope $\hat F^*(\lambda)$ with visited outer iterates overlaid (representative instance).
(b)~Active-branch structure across $\lambda$; switch points mark changes of the optimal bitstring.}
\label{fig:experiment9_grid_ab}
\end{figure}

Before presenting quantitative comparisons, we illustrate the bilevel optimization dynamics on a representative periodic instance. Figure~\ref{fig:experiment9_grid_ab} provides a multi-view diagnostic that makes both the outer landscape structure and the cost mechanism visible.

The periodic family is thereby particularly informative because its oscillatory weight functions create a multi-peaked outer landscape with frequent changes in the identity of the optimal bitstring. This ``switch-rich'' structure makes value probing both costly (each probe requires an inner re-solve) and directionally fragile (finite-difference probes at $\lambda \pm c$ may land on different active branches, corrupting the gradient estimate). 

\subsection{Systematic Budget Efficiency}
\label{sec:results_budget}
\begin{figure}[t]
\centering

\newcommand{\panelbox}[2]{%
\begin{minipage}[t]{0.49\columnwidth}
  \centering
  \textbf{(#1)}\par\vspace{0.25em}
  \includegraphics[width=\linewidth,height=0.95\linewidth,keepaspectratio]{figures/#2}
\end{minipage}%
}

\panelbox{a}{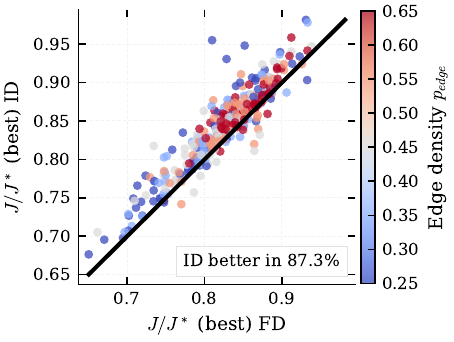}\hfill
\panelbox{b}{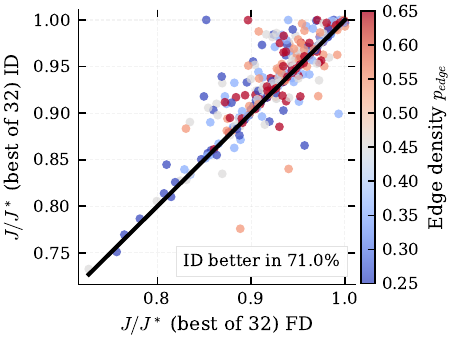}
\caption{\textbf{Bilevel optimization diagnostics for the periodic 1D family}
showing per-instance comparisons between CR-ID and central finite-difference value probing (FD) under a matched evaluation budget $B=1830$.
Each point corresponds to one problem instance ($N=300$), with color indicating the edge density $p_{\mathrm{edge}}$.
The black diagonal denotes equal performance.
Points above the diagonal indicate instances where ID outperforms FD.
(a) Best-so-far normalized expectation value $J/J^*$ at budget exhaustion.
(b) Best readout performance (best-of-32 samples) at the same budget.}
\label{fig:experiment9_grid_cd}
\end{figure}

To assess budget efficiency, we first compare CR-ID against the FD baseline on the periodic 1D benchmark under a matched objective-evaluation budget. The per-instance scatter in Figure~\ref{fig:experiment9_grid_cd} shows a consistent and systematic dominance of CR-ID in both the expectation-level best-so-far objective and the readout metric, aligning with the reduced overhead of probe-free outer-gradient estimation.

A quantitative summary of budget efficiency (AUC$_B$) for the 1D outer-control benchmark is provided in Table~\ref{tab:main_results}. CR-ID improves progress-per-budget across all tested response families (linear, quadratic, periodic), indicating a systematic benefit rather than a family-specific effect; the detailed values are deferred to the table. For the periodic family, complementary endpoint/readout/reliability metrics are reported separately in Table~\ref{tab:periodic_aux_metrics}, which also demonstrate the superiority of CR-ID in all metrics.

\begin{figure}[t]
\centering

\newcommand{\panelbox}[2]{%
\begin{minipage}[t]{0.49\columnwidth}
  \centering
  \textbf{(\lowercase{#1})}\par\vspace{0.25em}
  \includegraphics[width=\linewidth,height=0.95\linewidth,keepaspectratio]{figures/#2}
\end{minipage}%
}

\panelbox{a}{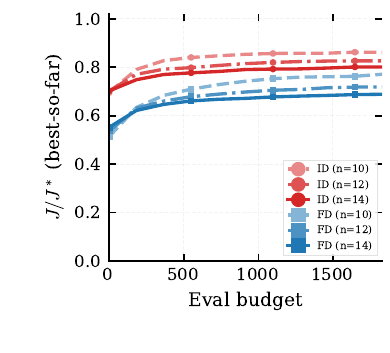}\hfill
\panelbox{b}{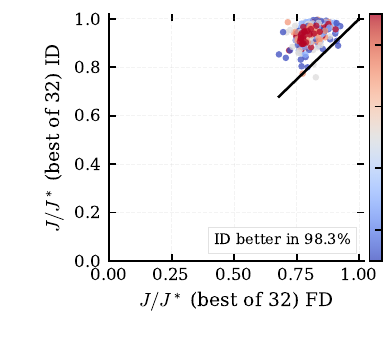}

\caption{\textbf{Budget efficiency in the multi-dimensional ($\boldsymbol{\lambda}$) setting under a matched evaluation budget.}
Here, the outer variable is \emph{edge-wise}, i.e., $\boldsymbol{\lambda}\in\mathbb{R}^{|E|}$ assigns one control parameter $\lambda_e$ per edge.
(a)~Best-so-far normalized objective $J/J^*$ versus cumulative energy evaluations.
(b)~Budget-matched best-of-$S$ readout $J/J^*$ across instances in the multi-dimensional setting, comparing ID and FD.
}

\label{fig:budget_efficiency_row1}
\end{figure}

Moving from 1D to multi-dimensional outer control further amplifies the advantage. In the edge-wise setting, where $\boldsymbol{\lambda}\in\mathbb{R}^{|E|}$ assigns a separate control $\lambda_e$ to each edge, CR-ID reaches higher best-so-far performance earlier over the evaluation budget and preserves this gain across problem sizes (Figure~\ref{fig:budget_efficiency_row1}). Under the matched budget, this translates into an approximately $14.4\%$ relative improvement in budget efficiency (AUC$_B$) over finite-difference value probing (Table~\ref{tab:md_outer_control}).

\begin{table*}[t]
\centering
\caption{\textbf{Budget efficiency summary for 1D outer control (AUC$_B$), budget-matched at $B=1830$.}
Values are mean $\pm$ std over $N=300$ instances.}
\label{tab:main_results}
\small
\begin{tabular}{lccc}
\toprule
\textbf{Family} & \textbf{CR-ID AUC$_B$} & \textbf{FD AUC$_B$} & \textbf{$\Delta$AUC$_B$} \\
\midrule
Linear & $0.7950 \pm 0.0672$ & $0.7653 \pm 0.0725$ & $+0.0297$ \\
Quadratic & $0.8193 \pm 0.0473$ & $0.7833 \pm 0.0512$ & $+0.0360$ \\
Periodic & $0.8061 \pm 0.0505$ & $0.7748 \pm 0.0562$ & $+0.0313$ \\
\bottomrule
\end{tabular}
\end{table*}

\begin{table*}[t]
\centering
\caption{\textbf{Auxiliary metrics for the periodic 1D outer-control benchmark, budget-matched at $B=1830$.}
Values are mean $\pm$ std over $N=300$ instances.}
\label{tab:periodic_aux_metrics}
\small
\begin{tabular}{lcc}
\toprule
\textbf{Metric} & \textbf{CR-ID} & \textbf{FD} \\
\midrule
Final best-so-far objective ($y(B)/J^*$) & $0.8404 \pm 0.0548$ & $0.8224 \pm 0.0538$ \\
Readout best (best-of-32) & $0.9339 \pm 0.0476$ & $0.9257 \pm 0.0462$ \\
Tail hit-rate $p_{\mathrm{good}}$ ($\varepsilon = 0.1$) & $0.3699 \pm 0.3025$ & $0.2607 \pm 0.2514$ \\
\bottomrule
\end{tabular}
\end{table*}

\begin{table*}[t]
\centering
\caption{Budget-matched results in the multi-dimensional outer-control setting (edge-wise $\boldsymbol{\lambda}\in\mathbb{R}^{|E|}$) at $B=1830$. Values are mean $\pm$ std.}
\label{tab:md_outer_control}
\begin{tabular}{lcc}
\hline
\textbf{Metric} & \textbf{CR-ID} & \textbf{FD} \\
\hline
$\mathrm{AUC}_B$ & $0.7742 \pm 0.0500$ & $0.6765 \pm 0.0445$ \\
Final best-so-far objective & $0.8272 \pm 0.0550$ & $0.7230 \pm 0.0532$ \\
Readout best-of-32 & $0.9336 \pm 0.0383$ & $0.8100 \pm 0.0407$ \\
Tail hit-rate ($\varepsilon = 0.1$) & $0.3192 \pm 0.2609$ & $0.0149 \pm 0.0987$ \\
\hline
\end{tabular}
\end{table*}

\paragraph{Trajectory dynamics.}
Figure~\ref{fig:budget_efficiency_row1}(a) further reveals qualitatively different convergence behavior between the two methods. CR-ID trajectories rise steeply in the early evaluation budget and plateau at high objective values across all tested problem sizes, indicating rapid convergence at a high solution quality. In contrast, FD trajectories increase more gradually throughout the budget window and remain in an upward trend at budget exhaustion, suggesting that convergence has not yet been reached. This pattern is consistent across problem sizes: CR-ID consistently attains its plateau before the matched budget $B$, whereas FD continues to improve slowly without saturating.

\paragraph{Where the budget goes.}
The observed trajectory separation has a direct structural explanation. Probe-based FD updates spend evaluations on value-function probes where each outer step requires three inner solves (center plus two perturbations), whereas CR-ID uses only the center evaluation and extracts the outer signal by reweighting correlators already measured for energy estimation. Under a fixed evaluation budget, CR-ID therefore completes approximately three times as many outer iterations as FD. This $3\times$ factor manifests visually as a horizontal compression of the CR-ID trajectory relative to FD: the same outer progress that CR-ID achieves by evaluation $b$ is reached by FD only around evaluation $3 \cdot b$. The AUC$_B$ improvements reported in Tables~\ref{tab:main_results} and~\ref{tab:md_outer_control} quantify this shift.

\paragraph{Endpoint similarity.}
Although CR-ID dominates budget-nor\-mal\-ized performance (AUC$_B$), both methods optimize the same outer landscape and would therefore be expected to reach similar final objective values given sufficient budget. The data support this interpretation: in the periodic 1D setting, the best-so-far objective at $B = 1830$ is $0.8404 \pm 0.0548$ (CR-ID) versus $0.8224 \pm 0.0538$ (FD), a gap that is notably smaller than the AUC$_B$ gap. Extrapolating from the FD trajectory slopes in Figure~\ref{fig:budget_efficiency_row1}(a), we expect FD to eventually plateau at a level comparable to CR-ID, but at a substantially larger evaluation cost. In other words, the advantage of CR-ID is not that it finds better solutions in principle, but that it finds equivalent good solutions \emph{earlier}—precisely the regime that matters when evaluation budgets are constrained, which is almost always the case.

\subsection{Architecture Comparison: VQE versus QAOA}
\label{sec:results_architecture}

\begin{table*}[t]
\centering
\caption{\textbf{VQE vs.\ QAOA (budget-matched at $B=1830$).}
Values are mean $\pm$ std over $N=300$. Readout uses best-of-32 samples.}
\label{tab:architecture_comparison}

\scriptsize
\setlength{\tabcolsep}{4pt}
\renewcommand{\arraystretch}{1.15}

\begin{tabular}{@{}p{0.42\columnwidth}ccc@{}}
\toprule
\textbf{Metric} & \textbf{VQE} & \textbf{QAOA} & \textbf{$\Delta$} \\
\midrule
Best-so-far $J/J^*$
& $0.8404 \pm 0.0548$
& $0.7029 \pm 0.0484$
& $+0.1375$ \\

AUC$_B$
& $0.7815 \pm 0.0496$
& $0.6815 \pm 0.0423$
& $+0.1000$ \\

Readout best $J/J^*$ (best-of-32)
& $0.9339 \pm 0.0476$
& $0.9031 \pm 0.0417$
& $+0.0308$ \\

Tail factor (readout / expectation)
& $1.1142 \pm 0.0700$
& $1.2880 \pm 0.0639$
& $-0.1738$ \\

Tail hit-rate $p_{\mathrm{good}}$ ($\varepsilon=0.1$)
& $0.3699 \pm 0.3025$
& $0.0476 \pm 0.0555$
& $+0.3223$ \\
\bottomrule
\end{tabular}
\end{table*}

\begin{figure}[t]
\centering
\begin{minipage}[t]{0.49\columnwidth}
  \centering
  \textbf{(a)}\par\vspace{0.25em}
  \includegraphics[width=\linewidth]{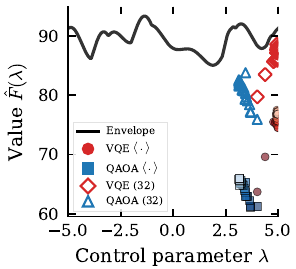}
\end{minipage}\hfill
\begin{minipage}[t]{0.49\columnwidth}
  \centering
  \textbf{(b)}\par\vspace{0.25em}
  \includegraphics[width=\linewidth]{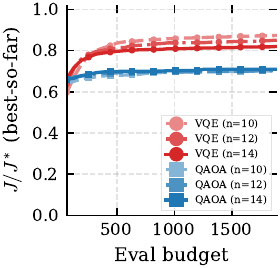}
\end{minipage}

\caption{\textbf{VQE versus QAOA under a matched objective-evaluation budget.}
(a) Representative instance: classical envelope $\widehat{F}(\lambda)$ (black) with the outer iterates $\{\lambda_t\}$ visited by VQE (red) and QAOA (blue); filled markers show expectation estimates and open markers indicate best-of-$S$ readout ($S=32$).
(b) Median best-so-far normalized objective $J/J^*$ versus cumulative objective evaluations for $n\in\{10,12,14\}$.}
\label{fig:vqe_vs_qaoa_row1}
\end{figure}

We next compare VQE and QAOA on the periodic benchmark under a matched objective-evaluation budget $B = 1830$ ($N = 300$). We use a hardware-efficient VQE ansatz with depth $L_{\mathrm{VQE}} = 2$ and QAOA with depth $p_{\mathrm{QAOA}} = 3$, and we evaluate readout-facing performance via best-of-$S$ sampling with $S=32$. For QAOA, the variational angles $(\boldsymbol{\gamma}, \boldsymbol{\beta})$ are warm-started using a linear ramp (adiabatic-inspired) initialization scheme---i.e., $\gamma_\ell$ is linearly increased and $\beta_\ell$ linearly decreased across layers~\cite{zhou2020quantum, chiew2023linearRampInit}. Moreover, we provide the full partial derivative $\partial_{\lambda} J(\boldsymbol{\theta}, \lambda)$, including the state-dependence term from Eq.~\eqref{eq:QAOA_decomp}, to isolate architectural effects from gradient-estimation artifacts. Figure~\ref{fig:vqe_vs_qaoa_row1} illustrates representative behavior and median budget trajectories, while Table~\ref{tab:architecture_comparison} reports the complete metric summary.

Across the evaluation budget, VQE consistently attains higher expectation-level best-so-far performance and correspondingly higher budget efficiency (AUC$_B$) than QAOA, indicating that the VQE architecture better leverages the clean correlator-reuse signal in this bilevel setting (see Figure~\ref{fig:vqe_vs_qaoa_row1}). When moving from expectation values to readout via best-of-$32$ sampling (see Table~\ref{tab:architecture_comparison}), the performance gap narrows because QAOA occasionally produces very high-quality bitstrings even when its expectation baseline is lower. This narrowing is driven by a stronger tail for QAOA, but it comes with substantially reduced reliability: near-optimal solutions occur more as rare events rather than as consistently sampled outcomes. The same picture is reinforced by per-shot sampling reliability, where VQE yields a much higher probability of drawing a solution within 10\% of the best achieved objective in a single shot, making its advantage decisive when readout budgets are limited or when consistent solution quality is required.

\section{Conclusion}
\label{sec:conclusion}
We have introduced a bilevel optimization framework for parametric diagonal cost Hamiltonians and developed correlator-reuse implicit differentiation (CR-ID) for efficient outer-loop tuning within this setting.

By leveraging the envelope theorem, CR-ID extracts outer gradients from correlators already measured during inner energy estimation—eliminating the $\approx\!3\times$ multiplicative overhead of derivative-free value probing. Our experiments demonstrate:

\begin{enumerate}
    \item \textbf{Systematic efficiency:} Under matched evaluation budget, CR-ID outperforms central finite-difference value probing across all tested 1D response families and multi-dimensional outer controls with up to 14.4\% relative improvement.

    \item \textbf{Architecture dependence:} The ``free'' nature of correlator-reuse is specific to VQE, where $\lambda$ affects only Hamiltonian coefficients. For QAOA, the state-dependence term~\eqref{eq:QAOA_decomp} creates a cost--bias trade-off that practitioners must navigate.
    
    \item \textbf{Operational metrics matter:} While VQE attains higher expectation-level performance than QAOA, best-of-32 readout reduces the gap substantially, although VQE remains in the lead. 

\end{enumerate}

\paragraph{Limitations.} Our evaluation is conducted at a modest system size ($n=12$) to enable classical diagnostics. The correlator-reuse argument exploits the structure of diagonal Hamiltonians; extending to non-diagonal Hamiltonians would require accounting for measurement grouping overhead. Additionally, the envelope identity is exact only at inner stationarity—our results demonstrate robustness under finite inner budgets, but stronger theoretical guarantees would require explicit characterization of the approximation error.

\paragraph{Outlook.} The bilevel formulation studied here captures a broad class of parametric quantum optimization problems. As quantum hardware improves, the measurement efficiency of CR-ID will become increasingly valuable. Future work could develop practical estimators for the QAOA state-dependence term and investigate connections to meta-learning in VQAs.

\section*{Acknowledgment}
This paper was partially funded by the German Federal Ministry of Education and Research through the funding program “quantum technologies -- from basic research to market” (contract number: 13N16196). Generative AI was utilized to generate sections of this Work, including text, tables, graphs, code, citations, etc.

\bibliographystyle{IEEEtran}
\bibliography{references}

\vspace{12pt}

\end{document}